\documentstyle[aps,prl,psfig,epsfig,multicol]{revtex}
\begin{document}
\def\be{\begin{equation}}
\def\ee{\end{equation}}
\def\bc{\begin{center}}
\def\ec{\end{center}}
\def\bea{\begin{eqnarray}}
\def\eea{\end{eqnarray}}
\draft
\title{Mean field solution of the Ising model on a
 Barab\'asi-Albert  network.}
\author{Ginestra Bianconi}
\address{Department of Physics, University of Notre Dame, Notre
Dame,Indiana 46556,USA } \maketitle
\begin{abstract}
The mean field solution of the Ising model on a Barab\'asi-Albert
scale-free network with ferromagnetic coupling between linked
spins is presented. The  critical temperature $T_c$ for the
ferromagnetic to paramagnetic phase transition ( Curie
temperature) is infinite and the effective critical temperature
for a finite size system increases as the logarithm of the system
size in agreement with recent numerical results of Aleksiejuk,
Holyst and Stauffer.
\end{abstract}

--------------------



 E-mail Address: gbiancon@nd.edu

--------------------

\begin{multicols}{2}
\narrowtext

The Ising model is the starting point for the study of second
order phase transitions and cooperative  phenomena. Traditionally
it has been solved exactly on periodic lattices in one and two
dimensions\cite{Yeomans} or in a Bethe lattice\cite{Thorpe}.
Recently attention has been addressed on the complex structure of
scale-free networks\cite{RMP} with power-law connectivity
distribution $P(k)\sim k^{-\gamma}$. These networks describe for
example biological systems where proteins are nodes and physical
protein interactions are the links\cite{Jeong01}. Moreover  they
constitute an intriguing substrate for the spreading of infectious
diseases\cite{Vespignani}, and for percolation
phenomena\cite{Havlin}. The Ising model has been applied to the
evolving scale-free network with $\gamma=3$ [the Barab\'asi-Albert
(BA) network\cite{BA}] by Aleksiejuk {\it et al.} (AHS)
\cite{Stauffer} and to random graph with 
arbitrary degree distribution \cite{DoroIS,Leone}.
By introducing a spin on each node of the network, they found that
ordering occurs for temperatures $T$ below the effective critical
temperature $T_c$, and that the nature of the phase transition is
very different from that on periodic lattices.

In this work  I show that a mean field solution of the Ising model
on the BA network can be treated as a Mattis model\cite{Mattis}
with the substitution $\xi_i\rightarrow k_i$ and that the scaling
of the effective critical temperature $T_c$ for ferromagnetic
transition increases as the logarithm of the system
 size in agreement with recent numerical calculations\cite{Stauffer}.

Let us consider a BA network of $N$ nodes. Starting from a small
number of nodes $n_0$   and  links $m_0$ ($n_0,m_0<<N$), the
network is constructed iteratively by the constant addition of
nodes with $m$ links. The new links are attached preferentially to
well connected nodes in such a way that at time $t_j$ the
probability $p_{ij}$ that the new node $j$ is linked to node $i$
with connectivity $k_i(t_j)$ is given by
\be
p_{i,j}=m \frac{k_i(t_j)}{\sum_{\alpha=1}^j k_{\alpha}}.
\label{p.eq} \ee If N is large we can approximate the total number
of edges present in the network at time $t_j$, given by the sum
$\sum_{\alpha=1}^j k_{\alpha}$ with $2mt_j$ and substituting into
$(\ref{p.eq})$ the dynamic solution for the
 connectivity \cite{BA}
\be
k_i(t)=m\sqrt{\frac{t}{t_i}}, \label{sol_BA.eq} \ee we obtain
\be
p_{i,j}=\frac{m}{2} \frac{1}{\sqrt{t_i t_j}}. \ee  The adjacency
elements of the network $\epsilon_{i,j}$ are equal one if there is
a link between node $i$ and $j$ and zero otherwise. Consequently
their mean over many copies of the BA network has  a tensor
structure: \bea [\epsilon_{i,j}]&=&p_{i,j}\nonumber \\
&=&\frac{m}{2} \frac{1}{\sqrt{t_i t_j}} \nonumber
\\ &=&\frac{1}{2mN}  k_i k_j. \label{e_ij} \eea

The Ising-type spins $s_i=\pm1$  placed on the nodes of a BA
network are subjected to the Hamiltonian
\be
H=-\sum_{i,j} J_{i,j} s_i s_i-\sum_i h_i s_i \ee with the local
magnetic field $h_i$ and the coupling
 $J_{i,j}$ non zero only for  nodes $i$ and $j$ connected by a link:
\be
J_{i,j}=J \epsilon_{i,j} \ee were $J>0$, and $\epsilon_{i,j}$ is
the adjacency matrix of the BA  network.

The exact solution of the Hamiltonian is given by
\be
<s_i>=<\tanh[\beta(\sum_j J_{i,j} s_j +h_i)]>. \label{sol_I.eq}
\ee

The mean field equation for the mean local magnetization $<s_i>$
is given by
\be
<s_i>=\tanh[\beta (J \sum_{j=1}^{N} [\epsilon_{i,j}] <s_j> +h_i)]
 \label{Is1.eq}
\ee where we  approximated $<\tanh(x)>$ with $\tanh(<x>)$ and we
 performed the mean of the adjacency matrix $[\epsilon_{i,j}]$
 over the different realization of the network. If we define
 $S$ by \bea S&=& \frac{1}{2mN}
\sum_{i=1}^{N} k_i <s_i> \eea  after substituting
$[\epsilon_{i,j}]$ by ($\ref{e_ij}$) we can rewrite
Eq.$(\ref{Is1.eq})$
\be
<s_i>=\tanh[\beta (J k_i S+h_i)].  \label{Is2.eq}\ee
 Multiplying both sides of the equation
by $k_i/2mN$ and summing over $i$  we can solve $(\ref{Is2.eq})$
self-consistently for S,
\be
S=\frac{1}{2 m N} \sum_{i=1}^{N} k_i \tanh[\beta (J k_i S +h_i)].
\label{Issol.eq} \ee

Thus the BA Ising model looks like a Mattis model\cite{Mattis}
with the substitution
\be
\xi_i \rightarrow k_i. \ee  This is interesting because the Mattis
model is a simple  solvable model of spin-glass \cite{Amit} and
suggest that in a spin-glass  on a scale-free network there will
be a noise term deriving both from the geometry of the system and
the random interactions.

 {\it The susceptibility--} The susceptibility of
 the order parameter $S$, describing the response to a uniform magnetic field $h$, is given by
\be
\frac{\partial S}{\partial h}=\frac{1}{2mN}\beta \frac{\sum_i
k_i(1-<s_i>^2)}{( 1-\frac{1}{2mN}\beta J \sum_i k_i^2
(1-<s_i>^2))}. \ee If we define the mean magnetization M as
\be
M=\frac{1}{N} \sum_i <s_i>, \ee the susceptibility
$\chi=\frac{\partial M}{\partial H}$ is given by
\be
\chi=\frac{1}{N}\sum_i(1-<s_i>^2)[\beta J k_i \frac{\partial
S}{\partial h}+\beta ] \ee

\begin{figure}
\centerline{\epsfxsize=3.5in \epsfbox{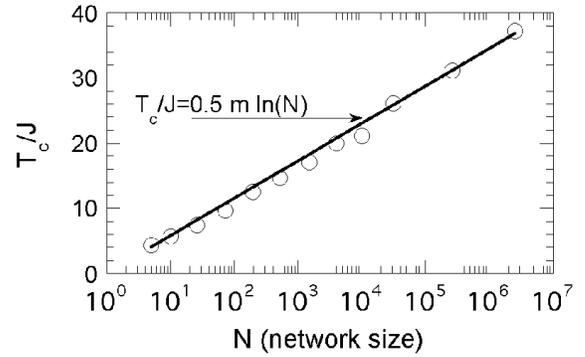}} \caption{The
analytical prediction for the ferromagnetic effective critical
temperature (solid line) is compared with the AHS simulations
(open circles) with $m=5$ as a function of the network size. }
\end{figure}
{\it The effective critical temperature--} Using $S$ as the order
parameter of the system, and its self-consistent definition
Eq.($\ref{Issol.eq}$),
 it is easy to find the effective critical temperature ($T_c$).
As $S\sim 0$ we can approximate the $\tanh(x)$ with $x$ and
$\beta$ with $\beta_c$ giving, \bea S&=&\frac{1}{2 m N} \int_1^{N}
dt' \beta_c J S k^2(t')\nonumber \\ &=&\frac{m}{2} J \beta_c S
\ln(N), \eea i.e.
\be
\frac{T_c}{J}=\frac{m}{2} \ln(N). \label{T_c.eq}\ee This shows
that for finite size system there is an effective critical
temperature that increases linearly with the interaction $J$ and
logarithmically with the number $N$ of the nodes of the network.
This result is in agreement with the numerical results of
AHS\cite{Stauffer} as it is shown in Fig. 1. In Fig. 1 we have
compared the analytical results Eq. $(\ref{T_c.eq})$ for the case
$m=5$ with the numerical results of AHS. However we know that a
finite system cannot have a phase transition, therefore the true
critical temperature goes to infinite in the thermodynamic limit
$N\rightarrow \infty$. This shows that in a scale-free network the
ordered phase is the only allowed phase in the thermodynamic
limit.

I am grateful to professor A.-L. Barab\'asi for useful discussion
and help. This work was supported by NSF.


\end{multicols}

\begin{references}

\bibitem{Yeomans}
J. M. Yeomans {'\it Statistical mechanics of phase transitions '}
(ed. Oxford Science Publications) (1992).
 \bibitem{Thorpe}
M. F. Thorpe, in {\it Excitations in Disordered Systems}, ed. M.
F. Thorpe (Plenum press, New York \& London,1982).
\bibitem{RMP}
  R. Albert and A.-L. Barab\'asi, {\it Rev. Mod. Phys.} {\bf 74},
  47 (2002).
  \bibitem{Jeong01}
 H. Jeong, S. P. Mason, A.-L. Barab\'asi and Z. N. Oltvai {\it Nature} {\bf 411},
  41 (2001).
\bibitem{Vespignani}
  R. Pastor-Satorras and A. Vespignani, {\it Phys. Rev. Lett.} {\bf 86},
  3200 (2001).
\bibitem{Havlin}
  R. Cohen, K. Erez, D. ben-Avraham and S. Havlin, {\it Phys. Rev. Lett.} {\bf 85},
  4626 (2000); R. Cohen, K. Erez, D. ben-Avraham and S. Havlin, {\it Phys. Rev. Lett.} {\bf 86},
  3682 (2001).
\bibitem{BA}
A.-L. Barab\'asi and  R. Albert, {\it Science} {\bf 286}, 509
(1999).
\bibitem{Stauffer}
A. Aleksiejuk, J. A. Holyst and D. Stauffer,  {\it Physica A} {\bf
310}, 260 (2002).
\bibitem{DoroIS}
S. N. Dorogovtsev, A. V. Goltsev and J. F. F. Mendes,
{\it Phys. Rev. E} {\bf 66}, 016104 (2002).
\bibitem{Leone}
M. Leone, A. Vazquez, A. Vespignani and R. Zecchina,
{\it Eur. Phys. J. B} {\bf 28},191 (2002).
\bibitem{Mattis}
D. C. Mattis {\it Phys. Lett.} {\bf 56 A}, 421 (1976).
\bibitem{Amit}
D. Amit, {\it Modeling brain function},(Cambridge University
Press, Cambridge, 1989)


\end{references}
\end{document}